\documentstyle[eqalign,epsf]{l-aa}
\begin{document}

\thesaurus{01         % A&A Section 1: Letter
     (02.07.2; % {Gravitational waves
      08.14.1; % Stars: neutron
      08.19.4)}% Supernovae:general

\title{High Neutron Star Birth Velocities and
Gravitational Radiation during Supernova Explosions}

\author{
S.N.Nazin\inst{1}
\and K.A.~Postnov \inst{1,2}}

\institute{
Sternberg Astronomical Institute, Moscow University,
                119899 Moscow, Russia
\and Faculty of Physics, Moscow University, 117234 Moscow, Russia
}
\date{Received ... 1996, accepted ..., 1996}
\maketitle

\begin{abstract}
Assuming the observed pulsar velocities to originate during
asymmetric collapse of stellar cores, we compute the amplitude of
gravitational waves emitted during type II and Ib supernova explosions
and their detection rate from within a distance of 30 Mpc.  At the
rms-level of advanced laser interferometers $h\approx 10^{-22}$ at
frequencies $300-1000$ Hz the expected rate is about 1 per year.


\end{abstract}

Supernova explosions are among the most violent events known in nature.
Supernovae type II and Ib are triggered by the gravitational collapse of
evolved degenerate cores of massive ($>10$ M$_\odot$) stars and leave
behind neutron stars or, possibly, black holes as a remnant. An
enormous amount of energy is released during the collapse (roughly, the
binding energy of a neutron star $\sim 0.15 M_\odot c^2$). Although
most of this energy is carried away by neutrinos, a part of it can
escape in the form of gravitational radiation provided that the
collapse is not purely symmetrical.  This collapse anisotropy
(expressed conveniently in terms of the fraction of the energy
released, $\epsilon M_\odot c^2$) was first introduced by Ozernoy
(1965) and Shklovskii (1970).

It required more than 20 years for the idea of the collapse anisotropy
to receive serious observational support.  First, the revision of the
pulsar distance scale led to increasing the observed pulsar velocities
by a factor of two up to values of about 400-500 km/s (Lyne \& Lorimer
1994). Observations of pulsars associated with young supernova remnants
revealed even higher pulsar velocities up to 900 km/s (Frail et al.
1994). Such high velocities of young pulsars could hardly be obtained
without additional recoil (``kick velocity'') during the supernova
explosion.  Next, a direct evidence for a kick velocity of at least
100 km/s has recently been obtained from observations of precessing
binary pulsar orbit in PSR J0045-7319 in the SMC (Kaspi et al. 1996).
The collapse asymmetry resulting in the kick velocity of a neutron star
at birth may be due to different reasons.  For example, recent
calculations of Burrows \& Hayes (1996) has shown the ability of
neutrino anisotropic emission to produce kick velocities of 400-500
km/s, as observed (see also Imshennik 1992; Bisnovatyi-Kogan 1993).

One of the consequences of Lyne and Lorimer's result was a recognition
that the kick velocity imparted to a neutron star at birth has
a power-law asymptotic form at high velocities (Lipunov, Postnov
\& Prokhorov 1996a,b). Using direct Monte-Carlo calculations
of binary star evolution (the so-called ``Scenario Machine''),
they found that the Lyne-Lorimer pulsar
transverse velocities are best reproduced assuming the
space (3-D) kick velocity distributed as
\begin{equation}
f(x)\propto \frac{x^{0.19}}{(1+x^{6.72})^{1/2}}
\label{LLkick}
\end{equation}
where $x=v/v_0$, $v_0=400$ km/s.
An important thing about having this distribution is that at high
velocities ($v > 500$ km/s) it goes practically as $v^{-3.17}$, that is
much slower than the maxwellian tail $\propto\exp(-v^2)$, which
sometimes is assumed for the kick velocity distribution (e.g. Portegies
Zwart \& Spreeuw 1996).

The anisotropic stellar collapse may be a source of gravitational waves
(GW).  To be detectable from the distance of Virgo cluster ($\sim 18$
Mpc) by future ground-based laser interferometric and bar detectors,
the GW energy emitted during a SN explosion should be about $\Delta
E=10^{-3} M_\odot c^2$ (Thorne 1995; Schutz 1996).  However, in
numerical calculations this energy has been found to be very tiny,
typically $10^{-9}-10^{-7}$ of the total energy output (see M\"uller
1996 for a review).  Unfortunately, all these calculations are still
far from being realistic considering enormous numerical difficulties
and uncertainties involved.

Nevertheless, the recognition of high additional velocities neutron
stars acquire at birth suggests a way to estimate $\epsilon$ from
observations, regardless of the unknown collapse anisotropy
mechanism(s).  Indeed, the kinetic energy of the neutron star motion,
$\Delta E=M_{NS} v^2/2$, where $M_{NS}$ is the neutron star mass, may
be considered as a lower limit to the energy emitted in GW. Assuming
$M_{NS}=1.4$ M$_\odot$, we obtain $\epsilon = 0.7\times (v/c)^2$. With
the distribution law (\ref{LLkick}) the fraction of high velocity
pulsars  is
\begin{equation}
P(>x)=\int_x^\infty f(\xi)d\xi \simeq
0.38\, x^{-2.17}
\label{P(y)}
\end{equation}
(the power-law asymptotics is valid for $x>2$).  For example, the
fraction of pulsars with $v>1000$ km/s is $\approx 0.05$. Of course,
there should exist a maximum cut-off velocity; it must be higher than
the maximum pulsar velocities observed, $\sim 2000$ km/s ($x>5$), but
its exact value only slightly changes the normalization coefficient.
For $\epsilon=10^{-4}$ ($v\approx 3200$ km/s)  we obtain $P(>8)\approx
0.004$, i.e.  every 250-th neutron star may be born with the high
velocity required.

The neutron star galactic birth-rate is determined by
that of massive ($>10$ M$_\odot$) stars, which is $\approx 1/25$
years assuming Salpeter mass function $f(M)dM\propto M^{-2.35}dM$
and the mean galactic star formation rate of 1 M$_\odot$ per year.
Using the relations $\epsilon=0.7 (v/c)^2$, $x=v/(400$km/s$)$ and
Eq. (\ref{P(y)}) we may write galactic birth rate of high-velocity
pulsars as
\begin{equation}
R=0.38/25\, x^{-2.17}\hbox{yr}^{-1}\approx 1.3\times 10^{-5}\epsilon_{-3}^{-1.085}
\hbox{yr}^{-1}
\label{gal_R}
\end{equation}
(here and below subscripts indicate
the quantities measured in units of the corresponding
power of ten, e.g. $\epsilon_{-3}\equiv \epsilon/10^{-3}$)

The characteristic amplitude of the GW burst from a SN located
at a distance $r$ is
(Thorne 1987; Schutz 1996)
\begin{equation}
h_c=5\times 10^{-22} \epsilon_{-3}^{1/2}
\left(\frac{1 \hbox{kHz}}{f_c}\right)^{1/2}
\left(\frac{18 \hbox{Mpc}}{r}\right)
\label{h_c}
\end{equation}
where $f_c$ is the characteristic frequency of the burst.  This level
should be detectable by the advanced LIGO interferometer.  The event
rate from the volume $V$ of the Universe is (e.g. Phinney 1991; see
also Lipunov, Postnov \& Prokhorov 1996b) ${\cal R}\approx
0.01\times(\hbox{galactic rate})\times (V/\hbox{Mpc}^3)$.  Using Eqs.
(\ref{gal_R}) and (\ref{h_c}) we find
\begin{equation}
\eqalign{
{\cal R}&\approx
(0.08\, \hbox{yr}^{-1})\,\epsilon_{-4}^{0.42}f_{1 \hbox{kHz}}^{-3/2}
h_{-22}^{-3} \cr
&\qquad\qquad\qquad\approx (1\, \hbox{yr}^{-1})\,\epsilon_{-3}^{0.42}f_{300
\hbox{Hz}}^{-3/2} h_{-22}^{-3}\,.
\cr}
\label{un_r}
\end{equation}
This means that we have chance to observe 1 GW-burst per year caused by
SN II and Ib explosions at a level of $h_{rms}=10^{-22}$ provided that
the energy is carried out at the frequency 300 Hz.

\begin{figure}
\epsfxsize=\hsize
\epsfbox{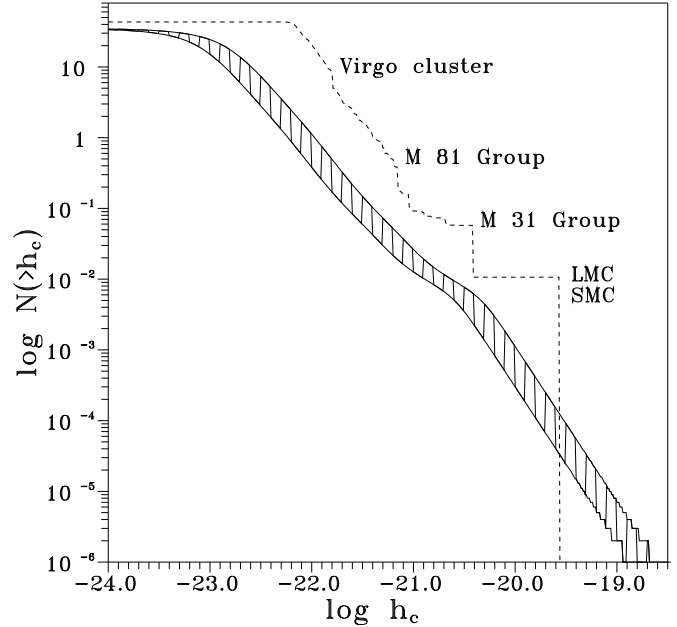}
\caption{The $\log N(>h_c) - \log h_c$ curve of the number of GW-bursts
caused by core collapse supernovae as seen by a GW-detector with an
rms-sensitivity $h_c$ at the unitary signal-to-noise ratio in 1-year
integration time. The supernova rate are calculated for the baryonic
matter distribution within the region of 30 Mpc from Sun according to
Tully's Nearby Galaxies Catalog (Tully 1988) with account for the
SN-rate dependence on the morphological type of the galaxies.  The
distribution of SN-generated GW-amplitudes is taken assuming
Lyne-Lorimer kick velocity distribution (see the text).  The hatched
region encompasses the assumed GW frequencies from 1 kHz (lower
boundary) to 300 Hz (upper boundary). The broken curve represents the
SN rate assuming constant GW energy fraction $\epsilon=10^{-4}$ (from
Lipunov et al. 1995) with contributions of local group of galaxies
indicated.}
\end{figure}

Since $h_c\propto \sqrt{\epsilon}\sim v$, the distribution of the
GW-bursts caused by the SN explosions from a given distance $r$ will
have the form (\ref{LLkick}). This enable us to compute the cumulative
distribution of the number of events with the amplitude higher than a
given one ($\log N(>h_c) - \log h_c$) using the realistic baryon
matter distribution within $\approx 30$ Mpc according to Tully's Nearby
Galaxies Catalog (Tully 1988) (see Lipunov et al. 1995 for more
detail). The result is presented in Fig. 1.  The hatched region
corresponds to GW-frequencies of the signal from 300 Hz (upper
boundary) to 1 kHz (lower boundary).  For comparison, we reproduce the
$\log N-\log h$ calculated for constant $\epsilon=10^{-4}$ (Lipunov et
al. 1995).  Clearly, the more realistic $\epsilon$-distribution yields
an order of magnitude smaller event rate because the rate of SN
explosions with higher $\epsilon$ strongly decreases.
The shape of the curve is smoothed by the $f(v)$ distribution.
The mean $\epsilon$ for Lyne-Lorimer velocity distribution
(\ref{LLkick}) is $\langle\epsilon\rangle
\propto \int v^2 f(v) dv \approx 4.4\times
10^{-6}$.

It is seen from the figure that a few events per year are expected at
the noise level of the advanced-LIGO rms-sensitivity $\sim 10^{-22}$.
At the initial laser interferometers sensitivity level $h_{rms}\approx
10^{-21}$ the expected SN rate is very low, $\sim 0.01$ per year.

In our analysis, we have not used any particular mechanism for the
collapse anisotropy and tried to rely on the observational properties
of pulsars which are reliable products of supernova explosions.  The
only assumption we used is that the pulsar kick velocities are entirely
due to the supernova explosion asymmetry.  Of course, not every core
collapse supernova may give rise to a pulsar. For example, some pulsar
studies suggest the galactic pulsar birth-rate $1/125-1/250$ yr$^{-1}$
(Lorimer et al. 1993), much smaller than the galactic SN II rate.
Additionally, some other mechanisms of the collapse anisotropy may
operate. This means that our results can be considered as a lower limit
and the actual detection rate may be higher. We conclude that insofar
as the observed pulsar velocities is a measure of the core collapse
anisotropy, gravitational radiation bursts from supernova explosions
should be detectable by the advanced laser interferometers and bar
detectors in 1-year integration.

The authors thank Prof. Vladimir Lipunov for helpful discussions.  The
work is supported by the INTAS grant No 93-3364, grant of Russian
Fund for Basic Research No 95-02-6053 and by the Center for
Cosmoparticle Physics ``COSMION'' (Moscow, Russia). The work of S.Nazin
was also supported by the grant of ISSEP No a96-1523.

\section*{Note added in proof}
In a recent paper, Hartman (1996) argued that
the observed transverse pulsar velocity distribution is reproduced
equally well assuming the Paczy\'nski (1990) distribution
for initial pulsar velocity $p(x)dx\propto (1+x^2)^{-2}$
with $x=v/(600 \hbox{km/s})$. The same analysis as for
our distribution (\ref{LLkick}) shows that the resulting
$\log N(>h_c) - \log h_c$ curve does not change appreciably
giving appoximately the same
$\langle\epsilon\rangle
\approx 4\times 10^{-6}$.

\end{document}